\documentclass[aps,prl,twocolumn,showpacs,superscriptaddress,floatfix]{revtex4}
\usepackage{graphicx}

\begin{document}

\title{Superdeformation   and   alpha  - cluster
structure in  $^{35}Cl$}

 \author{Abhijit Bisoi}

  \affiliation{Saha  Institute  of  Nuclear Physics, Bidhannagar,
Kolkata - 700064, INDIA}

  \author  {M.  Saha Sarkar}

  \thanks{corresponding author: maitrayee.sahasarkar@saha.ac.in}

\affiliation{Saha  Institute  of  Nuclear  Physics,  Bidhannagar,
Kolkata - 700064, INDIA}

 \author {S. Sarkar}

 \affiliation{Bengal Engineering and Science University, Shibpur,
Howrah - 711103, INDIA}

 \author{S. Ray}

 \affiliation{Saha  Institute  of  Nuclear  Physics, Bidhannagar,
Kolkata - 700064, INDIA}

 \author{ M. Roy Basu}

 \affiliation{University of Calcutta, Kolkata - 700009, INDIA}

 \author {Debasmita Kanjilal}

 \affiliation{Saha  Institute  of  Nuclear  Physics, Bidhannagar,
Kolkata - 700064, INDIA}

 \author {Somnath Nag}

 \affiliation  {Indian Institute of Technology, Kharagpur-721302,
INDIA}

 \author {K. Selva Kumar}

 \affiliation  {Indian Institute of Technology, Kharagpur-721302,
INDIA}

 \author {A. Goswami}

 \affiliation{Saha  Institute  of  Nuclear  Physics, Bidhannagar,
Kolkata - 700064, INDIA}

  \author {N. Madhavan}

  \affiliation  {Inter University Accelerator Centre, New Delhi -
110067, INDIA}

 \author {S. Muralithar}

 \affiliation  {Inter  University Accelerator Centre, New Delhi -
110067, INDIA}

 \author {R. K. Bhowmik}

  \affiliation  {Inter University Accelerator Centre, New Delhi -
110067, INDIA}

\date{\today}

\begin{abstract}

A  superdeformed (SD) band has been identified in a non - alpha -
conjugate nucleus  $^{35}Cl$.  It  crosses  the  negative  parity
ground  band  above  $11/2^-$  and becomes the yrast at $15/2^-$.
Lifetimes of all relevant states have been measured to follow the
evolution of collectivity. Enhanced B(E2), B(E1) values  as  well
as   energetics   provide   evidences  for  superdeformation  and
existence of parity doublet cluster structure in an odd-A nucleus
for the first time in A$\simeq$  40  region.  Large  scale  shell
model  calculations  assign  $(sd)^{16}(pf)^{3}$ as the origin of
these states. Calculated spectroscopic factors correlate  the  SD
states in $^{35}Cl$ to those in $^{36}Ar$.

\end{abstract}


\maketitle

The superdeformed bands observed in the even-even nuclei in upper
$sd$  shell  have  provided  favourable condition to describe the
collective rotation  microscopically  involving  the  cross-shell
correlations  \cite{1,2}.  Complementary descriptions in terms of
particle-hole excitations in  the  shell  model  \cite{2,4},  and
$\alpha$-clustering  configurations within various cluster models
\cite{clus, clus1, 28si} have  been  utilised  to  interpret  the
data.  Till  now  no such band has been observed in non - alpha -
conjugate odd-A, N$\neq$Z isotopes in this  region  \cite{clus2}.
If  a  nucleus  clusterizes into fragments of different charge to
mass ratios, the center of mass does not coincide with its center
of charge. As a result a sizeable static E1 moment may  arise  in
the   intrinsic   frame   \cite{clus3},   resulting   in  several
distinctive features in the spectra. Two adjacent opposite parity
deformed $\Delta I=2$ bands connected  by  strong  E2  intra-band
transitions  in  turn  are  connected  by  strong  E1  inter-band
transitions  \cite{clus3}  forming  an  apparent   $\Delta   I=1$
rotational  band  with  alternating  parity  states.  Since early
seventies   \cite{kimura,arima,nndc},   a   number   of   similar
alpha-cluster   bands  have  been  studied  extensively.  In  the
spectrum of $^{19}F$, cluster-model calculations have shown  that
coupling  of  a  proton  hole  in the $p$ shell coupled with four
nucleons in the $sd$ shell (a proton hole coupled  to  $^{20}Ne$)
gives  rise  to  alpha-cluster bands. The lowest alpha + $^{15}N$
parity partner bands built on $K^\pi= 1/2^+$  ground-state  band,
lowest  lying  famous  $K^\pi=  1/2^-$  at 110 keV and some other
bands lying above 5 MeV have been observed.  So  far  no  similar
clustering have been observed in odd A nuclei in the A$\simeq$ 40
region,  where  evidences  of  clustering have been manifested in
even-even nuclei through superdeformation.

According  to Ikeda \cite{ikeda}, in the spectra of light nuclei,
cluster like  configurations  would  appear  near  the  threshold
energy needed for breakup into proper sub-nuclei. For the nucleus
of  our  interest  $^{35}Cl$, the threshold energy \cite{mass} to
appear as a composite of $^{32}S$ and a triton  (t)  ($^{32}S+t$)
is  around  18 MeV. On the other hand, threshold for the decay of
the  composite  system  $^{35}Cl$  into  ($^{31}P$  +   $\alpha$)
clusters  is  around  6.5 MeV. The SD rotational band observed in
$^{36}Ar$ has been shown to have cluster structure.  So  one  may
expect  to  find deformed cluster bands in the excitation spectra
of $^{35}Cl$ also generated by  coupling  a  proton  hole  to  SD
states in $^{36}Ar$.

\begin{figure}
\vspace{-1cm}
\hspace{-1.  cm}
\includegraphics[height=10.cm,width=.8\columnwidth,angle=-90]{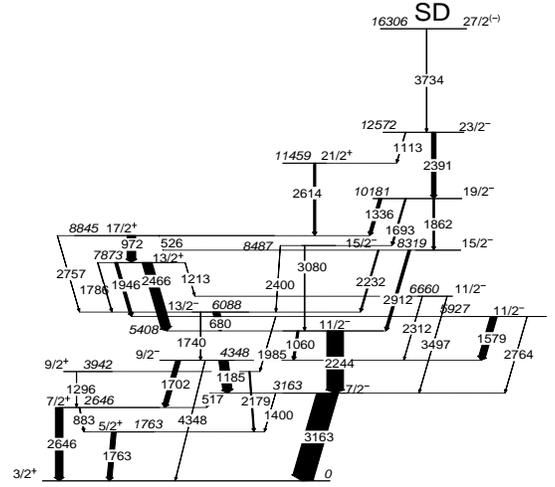}
%
\caption{\label{35cl}Partial level scheme of $^{35}Cl$.}
\end{figure}

In this letter, we report the observation of a superdeformed band
for  the  first  time in the odd A $^{35}Cl$ isotope. The reduced
transition   probabilities   for   all    relevant    transitions
depopulating  the  states  of  the yrast negative parity band and
related positive parity ones have been determined from  lifetimes
measured  in  the present experiment. Extracted B(E2)s and B(E1)s
provide important information to probe the remnant of  clustering
in  $^{35}Cl$.  Large  basis shell model (LBSM) calculations have
been done to understand the evolution of collectivity along  this
band.  The  predicted \cite{clus} negative parity partner band of
the SD band in $^{36}Ar$ has  also  been  reproduced  within  LBSM
calculations  for the first time.  Spectroscopic factors
have been used to correlate the SD states  in  $^{35}Cl$  to  the
cluster  states  in  $^{36}Ar$  to  establish  the persistence of
alpha-clustering features.

High  spin  states  of  $^{35}Cl$  have  been  populated  through
$^{12}C$($^{28}Si$,$\alpha$p)$^{35}Cl$ reaction  in  the  inverse
kinematics  with  an  110  MeV $^{28}Si$ beam at Inter University
Accelerator Center (IUAC), New Delhi. The target was $^{12}C$ (50
$\mu$g/$cm^{2}$) evaporated on 18  mg/$cm^{2}$  Au  backing.  The
$\gamma$-$\gamma$ coincidence measurement has been done using the
multi  detector  array  of  thirteen  Compton  suppressed  clover
detectors (INGA setup). The relevant details of the  experimental
setup  have  been  discussed  in  \cite{inga}. The detectors were
placed at  148$^\circ$  (4),  123$^\circ$  (2),  90$^\circ$  (4),
57$^\circ$  (2)  and  32$^\circ$  (1). The experimental data have
been sorted into  angle  dependent  symmetric  ($90^o  ~\rm  {vs}
~90^o$)  and  asymmetric  $\gamma$-$\gamma$  matrices  to get the
information about the gamma intensities,  DCO  ratios  and  level
lifetimes of this band.

\begin{figure}
\vspace{2cm}
\includegraphics[width=.9\columnwidth]{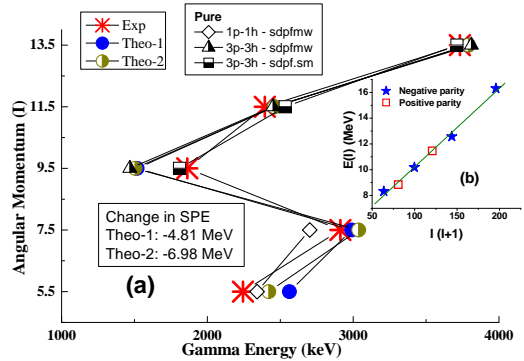}
\vspace{-3.cm}

\caption{\label{backbend}  (a) Comparison   of   experimental   and
theoretical backbending plots for the yrast negative parity  band
in  $^{35}Cl$.  The figure shows the results of calculations with
fixed n$\hbar\omega$ and  mixed  n$\hbar\omega$
truncations             {\it             viz.},             Theo1
:[$(1d_{5/2})^{12}(2s_{1/2}1d_{3/2})^6(pf)^{1}             \oplus
(1d_{5/2})^{12}(2s_{1/2}1d_{3/2})^{4}(pf)^{3}$];           Theo2:
[$(sd)^{18}(pf)^1                                          \oplus
(1d_{5/2})^{12}(2s_{1/2}1d_{3/2})^4(pf)^3$].    The   amount   of
depression of the single particle energies (SPE) of $pf$ orbitals
in each truncation scheme  is  also  mentioned.  (b)The  energies
(E(I))  of  $15/2^-$,  $17/2^+$,  $19/2^-$, $21/2^+$,$23/2^-$ and
$27/2^-$ states are plotted as function of I(I+1)}

\end{figure}

In  our  earlier  work  \cite{npa}, we have discussed about a few
totally shifted gamma rays in the  spectra  emitted  from  states
having  lifetimes  shorter  than the characteristic stopping time
($\simeq 10^{-13}$ s) of $^{35}Cl$ recoils in gold (Au)  backing.
Analysis of the present data confirmed \cite{vedova} that four of
them  (2912  keV,  1862 keV, 2391 keV and 3734 keV) belong to the
same  sequence  (marked  as  SD  in  Fig.~\ref{35cl}),  which  is
connected  to  the  other states in $^{35}Cl$ through 1113, 1336,
2232 and 1693 keV  transitions  (Fig.  ~\ref{35cl}).  The  lowest
state of this sequence 5408 keV ($11/2^-$) is already known to be
connected  to  the  3163  keV ($7/2^-$) through a strong 2244 keV
transition \cite{nndc}. This sequence  from  3163  keV  ($7/2^-$)
state  to  16306 keV (27/2$^-$) has been established as the yrast
negative parity band. The three topmost transitions in  the  band
show  nearly  perfect  rotational  behaviour by the almost linear
increase in angular momentum with gamma - ray energy  (rotational
frequency)  (Fig.  \ref{backbend}).  The plot shows a sharp break
between $15/2^-$ to $19/2^-$ indicating a  crossing  between  two
weakly   interacting   bands  of  different  configurations.  The
kinematic moment of inertia (KMOI) for the top three  transitions
in  yrast  band  in  $^{35}Cl$ scaled by the mass factor compares
very well with those for SD bands in $^{36}Ar$ and $^{40}Ca$. The
average KMOI, $\simeq$ 8 $\hbar^2/MeV$ also  compares  well  with
the   newly  found  candidate  SD  band  in  $^{28}Si$  ($\simeq$
6$\hbar^2/MeV$) \cite{28si}. More dramatically, the inset of Fig.
\ref{backbend} shows that energies of alternating parity  states,
$15/2^-$,  $17/2^+$,  $19/2^-$,  $21/2^+$,  $23/2^-$ and $27/2^-$
follow a linear relation  to  the  I(I+1)  values  exhibiting  an
apparent  rotational  band structure, which characterises nuclear
molecular structure \cite{clus3}.

\begin{figure}
\vspace{2cm}
\hspace{-1.cm}
\includegraphics[height=5cm,width=1.\columnwidth]{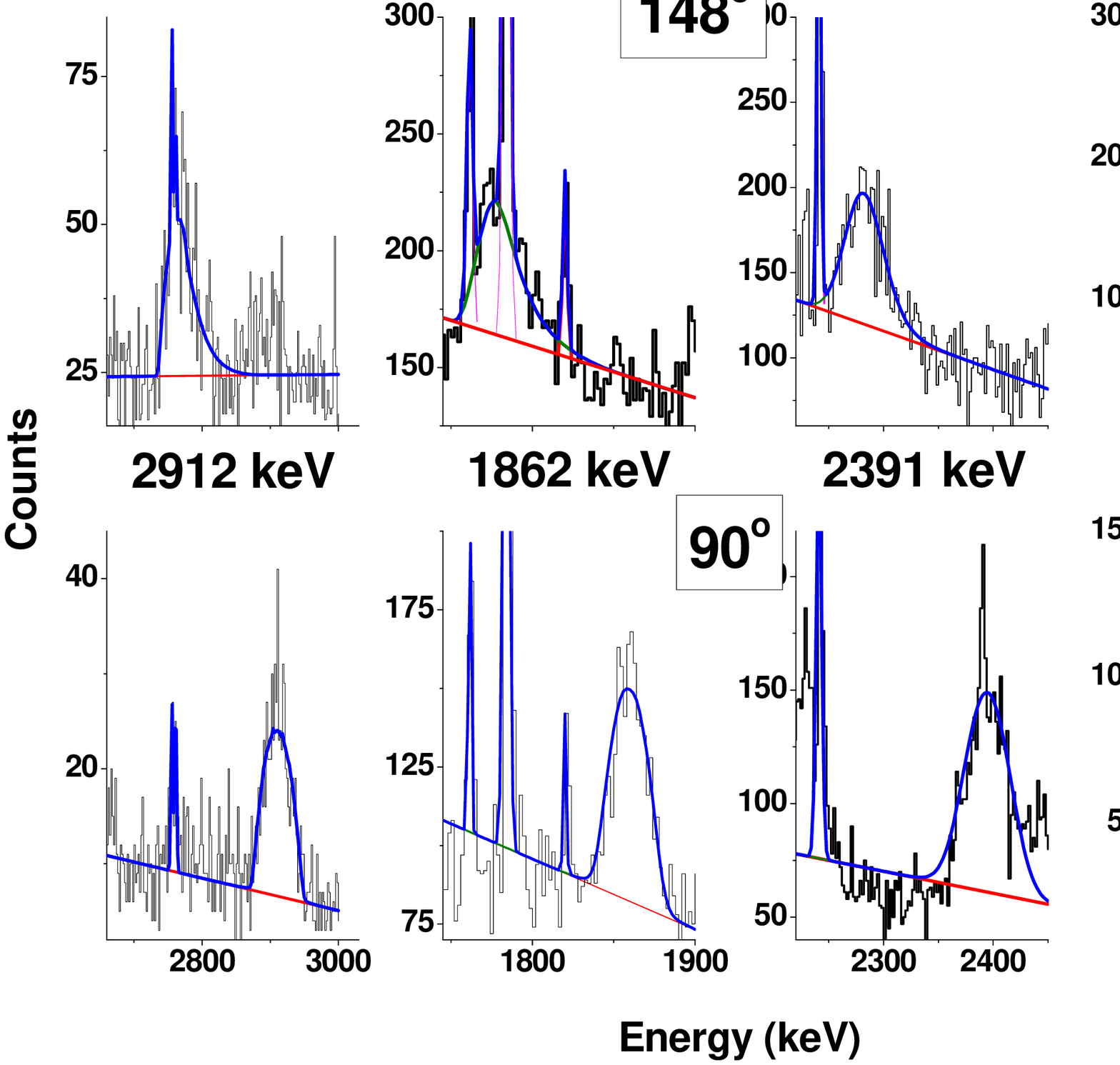}
\vspace{-2.5cm}

\caption{\label{lineshape}    The    Doppler-shift    attenuation
lineshape  analyses   for   different   relevant   gamma   rays.}

\end{figure}

\begin{figure}
\vspace{2cm}
\includegraphics[height=5cm,width=\columnwidth]{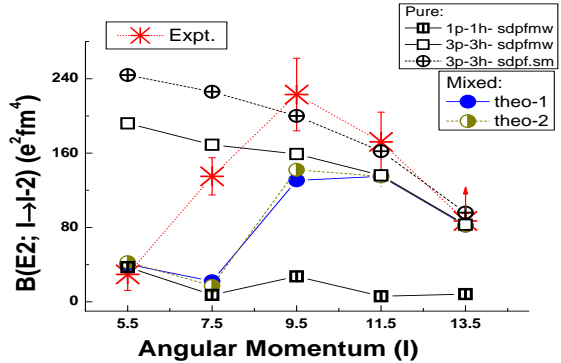}
\vspace{-2.5cm}

\caption{\label{be2}  $B(E2;I_i  \rightarrow I_f)$ values for the
yrast band in $^{35}Cl$ compared with the results of shell  model
calculations.  }

\end{figure}

Lineshapes  of  all  gamma rays in the yrast sequence (except the
2244 keV gamma), along with the 2614 keV gamma ray connecting the
$17/2^+$ state to $21/2^+$, have been analysed to  determine  the
lifetimes  of  the  corresponding states. Due to the large recoil
velocity in inverse kinematics and high energies of the  relevant
gammas,  special  care  has  been  taken  to  choose  the  gating
transitions such that the slow feeding components are eliminated.
Gating from top  could  not  be  done  due  to  poor  statistics.
However,  while choosing a suitable gating transition from below,
special care has been taken  to  eliminate  the  contribution  of
strong gamma peaks close to the peaks of interest. The spectra at
mean  angles of $148^o$ and $90^o$ relative to the beam axis were
simultaneously fitted  (Fig  \ref{lineshape})  using  a  modified
version  of the LINESHAPE \cite{line1, line2} code which included
corrections for the broad initial  recoil  momentum  distribution
produced   by  the  $\alpha$-particle  evaporation.  The  initial
momentum distribution of $^{35}Cl$ recoils has been obtained from
statistical model code PACE4  \cite{pace4}.  Lineshapes  of  four
gamma  transitions  (3734,  2391,  1862 and 2912 keV) were fitted
simultaneously as  members  of  a  single  band.  The  rotational
cascade  side-feeding  has  been  considered, assuming 100\% side
feeding into  the  top  of  the  band.  During  each  line  shape
simulation,   the   background  parameters,  intensities  of  the
contaminant  peaks,  and  side-feeding  quadrupole  moments  were
allowed  to  vary, and the best fit set was obtained by using the
$\chi^2$ minimization. The lineshape of the 1336  keV  transition
has  also  been  fitted  to  get  an  independent estimate of the
lifetime of the 10181 keV level.  Using  the  relevant  branching
ratios  \cite{nndc},  experimental  B(E2)s  were  determined from
these  lifetimes  (Fig.   \ref{be2},   Table   \ref{be2t}).   The
collectivity  in  this  negative  parity band evolves from single
particle excitations (B(E2) $\simeq$ 5 W.u.) at lower spins, to a
set of highly deformed SD states (B(E2) $\simeq$ 20-33  W.u.)  in
between, finally terminating at a state at $27/2^-$ with moderate
deformation  (B(E2) $\simeq$ 13 W.u.). The gamma transition (2614
keV) connecting the positive parity states  also  has  comparable
B(E2)  ($\simeq 16$ W.u.) The decay-out transitions (1336 keV and
1113 keV) have relatively large B(E1) values, {\it viz.}, 2.9 and
2.6  $\times  10^{-3}$  W.u.,  respectively.  In  $^{36}Ar$   and
$^{40}Ca$,  with  4p-4h  and  8p-8h excitations in the $pf$ (N=3)
shell, the deformation ($\beta_2$) in  the  SD  bands  were  0.45
\cite{2}  and $\simeq$ 0.59 \cite{1}, respectively. In $^{35}Cl$,
3p-3h excitations (discussed below) give  rise  to  SD  structure
with comparable KMOI but relatively smaller deformation ($\simeq$
0.37)  (Table  \ref{be2t})  similar to the observation in heavier
nuclei where the occupation numbers of high-N orbitals have  been
found to characterise SD bands.

\begin{table}
\caption{\label{be2t}  Experimental level lifetimes, experimental
and theoretical (Theo2) reduced transition  probabilities  (B(L))
for  different  transitions,  corresponding transition quadrupole
moments,  Q$_t$  (eb),  together  with   quadrupole   deformation
($\beta_2$)  and  X  (major  to minor axis ratio of an ellipsoid)
deduced \cite{sd} from $Q_t$s are tabulated. The units of  B(E1),
B(E2)   and   B(M1)   are   $  10^{-3}  e^2fm^2$,  $e^2fm^4$  and
$10^{-2}\mu_N^2$ , respectively. The effective charges are $e_p$
=  1.5e,  $e_n$  =  0.5e.  For  B(M1)  calculations,  free single
particle g factors have been used.}

\begin{tabular} {cccccccc}
\hline
$I_i^\pi $&$\tau_{mean}$ &$E_{\gamma}$ &\multispan{2} \hfil B(L)\hfil&Q$_t$&$\beta_2$&X \\
&fs&keV &Exp.&Theo2&eb \\
\hline
11/2$^-$& 400(100)&2244 (E2)&  29(8)&43&0.31(4)&0.15&1.16\\
& \cite{lornie}\\
15/2$^-$&18(1) &2912 (E2)& 135(10)&17&0.65(2)&0.30&1.32\\
&&2232(M1)&$<$10&1.3\\
19/2$^-$&61(4)&1862 (E2)&224(31)&142&0.82(6)&0.37&1.41\\
& 61(7) &1336 (E1)&1.98(2)&1.03\\
& &1693 (E2)&148(41)&15\\
23/2$^-$&48(6)&2391 (E2)&172(22)&135&0.71(5)&0.32&1.35 \\
&&1113 (E1) &1.8(2)& 1.22\\
27/2$^-$&$< 13$&3734 (E2)& $>87$&82&$>$0.5&0.24&1.25\\
21/2$^+$&$62(3)$&2614 (E2)& 108(5)&67&0.56(2)&0.27&1.28\\
\hline
\end{tabular}
\end{table}

Large  basis shell model (LBSM) calculations have been done using
the SDPFMW Hamiltonian \cite{6} (as referred to within the OXBASH
code package \cite{OXB}). The  SDPF.SM  \cite{1,2,4}  Hamiltonian
suitable  for  fixed  n$\hbar\omega$ \cite{4} excitation has also
been used. The valence space consists of  full  $sd-pf$  orbitals
for  both protons and neutrons above the $^{16}O$ inert core (see
Ref. \cite{npa} for  details).  The  negative  (positive)  parity
spectra  have been calculated with pure $\hbar \omega$ =1 (0) and
3 (2) excitations (Fig. \ref{backbend}). The  experimental  gamma
energies  ($E_\gamma=E(I)-  E(I-2)$) and corresponding B(E2)s for
lowest state (I=11/2$^-$) and the upper three states  (I=19/2$^-$
to  27/2$^-$)  agree  reasonably  well with the calculated values
with 1p-1h and 3p-3h excitations, respectively. The results  with
SDPF.SM  exhibit  better agreement with the experimental data for
spins 19/2$^-$ to  27/2$^-$  (Figs.  \ref{backbend},  \ref{be2}).
However,  for  I=$15/2^-$,  although the gamma energy agrees well
with 1p-1h results, the experimental B(E2) matches better to  the
theoretical B(E2) values from 3p-3h calculations (Fig.\ref{be2}).
The  reduction  in  experimental  B(E2) at $27/2^-$ is reproduced
well by theory.  This  decrease  indicates  band  termination  at
$27/2^-$,   consistent   with   a  proton  hole  coupled  to  the
terminating spin (16$^+$) in the superdeformed band in  $^{36}Ar$
nucleus \cite{2,4}.

\begin{table}
\caption{\label{wavefunc}  Comparison  between  wavefunctions  of
different states obtained  from  empirical  fit  and  theoretical
calculations.   The  3p-3h  components  can  be  determined  from
normalisation condition.}

\begin{tabular} {cccccc}
\hline
$E_x$& $I_i $& $E_\gamma^{decay}$&\multispan{3} \hfil Wavefunction (1p-1h)\hfil \\
&&&Emp.&Theo1&Theo2\\
\hline
10181&19/2$_1$&1862&0&-&0\\
8319& 15/2$_1$&2912& 9&13 & 8\\
5408& 11/2$_1$&2244& 16&68 & 76\\
3163& 7/2$_1$&3163& 94&75&80\\
\hline
\end{tabular}
\end{table}

Configuration mixing between 1p-1h, 3p-3h configurations has been
included  for  further  improvement of the results. The set Theo1
has inert $1d_{5/2}$ in 1p-1h excitation, whereas in Theo2,  this
orbital  was  active  (Fig.  \ref{backbend}).  Inclusion of 5p-5h
configurations has been found to be insignificant.  It  has  been
shown  earlier that to reproduce the experimental data, the $sd -
pf$ shell gap has to be decreased depending upon  the  particular
truncation  scheme  involved  \cite{npa,mss}. In the experimental
spectra (Fig.\ref{35cl}), two close lying $15/2^-$ states (energy
difference is $\simeq$ 169 keV) are  seen.  The  single  particle
energies  (SPE)  of the $pf$ orbitals have been shifted downwards
to reproduce  the  splitting  between  the  two  lowest  $15/2^-$
states.  Results  from  mixed  calculations  show  improvement in
reproducing  the   gamma   energy   of   $15/2^-$   state   (Fig.
\ref{backbend})  deteriorating  the  agreement  for  B(E2;$15/2^-
\rightarrow 11/2^-$) value. The reduced transition  probabilities
of  decay-out E1 transitions (1336, 1113 keV) are reproduced well
(Table  \ref{be2t}).  Shell  model  calculations  reproduce   the
transition  energy and B(E2)of the 2614 keV transition reasonably
well. The highest  limit  of  experimental  B(M1)  for  2232  keV
transition  is  larger  than  the  predicted  value. However, the
calculated values of B(E2; $19/2^-_1 \rightarrow 15/2^-_2$) (1693
keV) and B(E2; $15/2^-_1 \rightarrow 11/2^-_1$) (2912  keV)  both
are   severely   underpredicted  (Table  \ref{be2t}),  indicating
inadequacy of the configuration mixed calculations. In $^{36}Ar$,
$^{40}Ca$ also \cite{2,4}, the calculations failed  to  reproduce
the  transition  probabilities  for  the  states  where different
configurations interact to their maximum.

A  simple  phenomenological approach \cite{fortune30mg} using two
level mixing between pure 3p-3h and 1p-1h states have  been  used
to  determine  the extent of configuration mixing existing in the
states near the band crossing. In this calculation  the  $19/2^-$
state has been assumed to be a 3p-3h state (100\%) with no mixing
from  1p-1h  (0\%). Utilising the transition matrix elements from
pure  1p-1h  and  3p-3h  LBSM  calculations,  the  experimentally
observed  B(E2;  I  $\rightarrow$ I-2) values for the transitions
with I$< 19/2^-$ have been reproduced considering the  the  mixing
coefficients  of  two  component  wavefunction  for each state as
variables (Table \ref{wavefunc}).  The  shell  model  predictions
deviate   from  the  phenomenologically  determined  wavefunction
structure only for the $11/2^-$ state. It is  evident  that  this
deviation  leads to the large difference between the experimental
and predicted (Theo1 and Theo2) B(E2) for the $15/2^- \rightarrow
11/2^-$ transition.

\begin{figure}
\vspace{2cm}
\includegraphics[height=6cm,width=1.1\columnwidth]{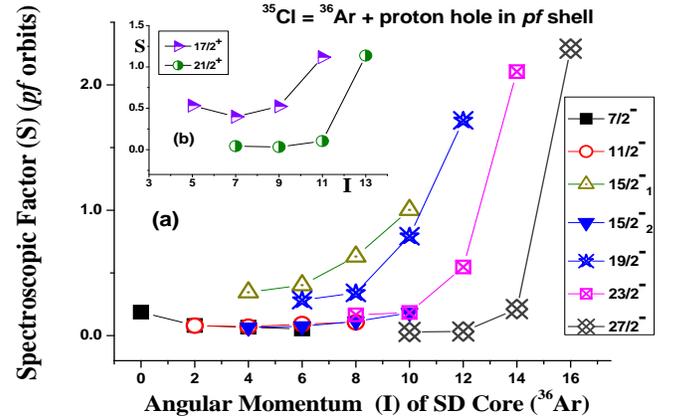}
\vspace{-3.cm}

\caption{\label{specfac} Calculated spectroscopic factors (Theo1)
for  the  (a)  negative  parity odd-spin  SD  states  in $^{35}Cl$ and (b)
positive parity partner even spin states to estimate  the  contribution  of
the  positive parity SD states and their negative parity partners
of the core nucleus, $^{36}Ar$.}

\end{figure}

The  SD  band  in  $^{36}Ar$ has been shown to originate from the
band crossing  of  ($^{32}S$($I^\pi$  =  $0^+-8^+$)  +  $\alpha$)
cluster  bands  \cite{clus}.  The  existence of a negative-parity
partner band of the SD band is also predicted  \cite{clus}.  Even
in  shell model picture, a negative parity partner of the SD band
in $^{36}Ar$ has been  obtained  in  terms  of  3p-3h  excitation
($(sd)^{17}  (pf)^3$)  in the $pf$ shell. So far, it has not been
verified experimentally. The parentage of the negative parity  SD
states  in  $^{35}Cl$  in  terms  of a proton hole coupled to the
alpha cluster SD states in $^{36}Ar$  core,  have  been  obtained
from    calculated    spectroscopic    factors    (Theo1)   (Fig.
\ref{specfac}). The $15/2^-$ to $27/2^-$ states in $^{35}Cl$  are
generated primarily from the cluster SD states, 10$^+$ to $16^+$,
whereas,  the  $17/2^+$ and $21/2^+$ arise predominantly from the
11$^-$ and $13^-$ states in the negative parity partner  band  in
$^{36}Ar$. The observed $17/2^+$ and $21/2^+$ states in $^{35}Cl$
therefore provide indirect experimental evidence in favour of the
existence of a negative-parity partner band of  the  SD  band  in
$^{36}Ar$ as predicted in Ref \cite{clus} and present work.

 Using  the  weak  coupling  model  \cite{arima}, the interaction
energy of (SD states in $^{36}Ar$ + $\pi  1f_{7/2}$  hole)  which
gives rise to the SD states in $^{35}Cl$ has been calculated. The
E($\pi  1f_{7/2})_{hole}$  has been estimated from the excitation
energy of the  7/2$^-$  state  in  $^{35}Cl$  at  3163  keV.  For
example,  $V_{int}  (15/2^-$  in $^{35}Cl$) = [\{BE $(10^+$ SD in
$^{36}Ar$)-E(hole  in   $1f_{7/2}$)\}-   BE   ($15/2^-$   SD   in
$^{35}Cl$)]  These  estimations show that to generate $15/2^-$ to
$27/2^-$ SD states in $^{35}Cl$, by coupling a proton  hole  with
$10^+$ to $16^+$ SD states in $^{36}Ar$, the interaction energies
\cite{mass}  are  915  keV,  175  keV,  -382  keV  and  -714 keV,
respectively. These are small  compared  to  a  few  MeV  average
particle-particle  interaction,  and  in  two  of  the  cases are
repulsive, which is favourable for cluster  structure  formation.
The unequal masses of the underlying clusters give rise to dipole
degree  of  freedom  leading  to  the  formation  of  an apparent
rotational   band   containing   alternating   parity    sequence
(Fig.\ref{backbend}) connected by strong E1 transitions. These E1
transitions  are  stronger than 2$\times 10^{-3}$ W.u. indicating
that  the  corresponding  positive  parity  states  are   doublet
partners \cite{clus3}.

To  summarize,  a  superdeformed  rotational  band  with  average
transition quadrupole moment $Q_t$ $\simeq 0.75$ eb and kinematic
moments of inertia $\simeq 8 \hbar^2/MeV$ has been identified for
the first time in odd $^{35}$Cl which terminates at $27/2^-$. The
energies  of  alternating  parity  states,  $15/2^-$,   $17/2^+$,
$19/2^-$,  $21/2^+$,  $23/2^-$  and  $27/2^-$ connected by strong
B(E1)s ($>  2\times  10^{-3}$  W.u.)  and  strong  cross-over  E2
transitions,  follow  a  linear  relation  to  the  I(I+1) values
providing a strong evidence in favour  of  cluster  structure  of
these states. Results of LBSM calculations have shown that the SD
band     has     been     generated    by    3p-3h    excitations
($(sd)^{16}(pf)^{3}$configuration). A  simple  two  level  mixing
calculation  has  been done to extract the wavefunction structure
of these mixed states from the experimental  B(E2)  values.  Weak
coupling  estimation  in  terms  of  $^{36}Ar$  and a proton hole
clearly identifies the origin of each SD state. The spectroscopic
factors are calculated to estimate parentage of these  SD  states
and their partners in terms of a proton hole coupled to the alpha
cluster SD states in $^{36}Ar$ core.The observed $17/2^+$ and $21/2^+$ 
states in $^{35}Cl$ therefore  provide  indirect experimental evidence
in favour of the existence of a negative-parity partner band of  
the  SD  band  in $^{36}Ar$ as predicted in Ref \cite{clus} and 
present work.

The   authors   acknowledge   the   help   from  all  other  INGA
collaborators and the Pelletron staff of IUAC for  their  sincere
help  and  cooperation.  Special thank is due to Pradipta Das for
his technical help for target preparation.  Discussions  with  P.
Banerjee  and  A  K  Singh  during  analysis of the DSAM data are
gratefully acknowledged. One  of  the  authors  (A.B.)  has  been
financially  supported  by  Council  of Scientific and Industrial
Research     (CSIR),      India,      under      contract      No
09/489(0068)/2009-EMR-1.

\end{document}